\documentclass{svjour3}

\usepackage{amsmath,amssymb}
\usepackage{hyperref}

\usepackage[amsmath, thmmarks]{ntheorem}
{\theoremstyle{nonumberplain}
\theoremsymbol{\ensuremath{\Box}}
\newtheorem{pf}{Proof}}

\newtheorem{thm}{Theorem}
\newtheorem{lem}{Lemma}
\newtheorem{cll}{Corollary}
\newtheorem{df}{Definition}
\newtheorem{rk}{Remark}

\title{Generalized Hamming weights of three classes of linear codes}

\author{Gaopeng Jian    
}

\institute{Gaopeng Jian \at
         School of Mathematical Sciences, Peking University, Beijing, P.R.China; \\
              \email{gpjian@pku.edu.cn}           
}

\date{}

\begin{document}

\maketitle

\begin{abstract}
The generalized Hamming weights  of a linear code have been extensively studied since Wei first use them to characterize the cryptography performance of a linear code over the wire-tap channel of type II. 
In this paper, we investigate the generalized Hamming weights of three classes of linear codes constructed  through defining
sets and determine them partly for some cases.
Particularly, in the semiprimitive case we solve an problem left in Yang et al. (2015) \cite{yang2015generalized}.
\end{abstract}

\section{Introduction}
Throughout this paper let $p$ be an odd prime and $q=p^m$ for some positive integer $m$.
Let $n_1$ be a positive integer coprime to $p$ and without loss of generality we assume that $m$ is the least positive integer such that $p^m \equiv 1 \pmod{n_1}$.
Denote by  $\mathbb{F}_p$ (or $\mathbb{F}_q$) the
finite field with $p$ (or $q$) elements.
Let $\alpha$ be a fixed primitive element of  $\mathbb{F}_q$.
Let $N=\frac{q-1}{n_1}$,  $N_1=gcd(N, \frac{q-1}{p-1}) $, $N_2=lcm(N, \frac{q-1}{p-1}) $ and $\theta=\alpha^N$.
Let $\text{Tr}$ denote the trace function from $\mathbb{F}_q$ to $\mathbb{F}_p$.  
An $[n,k,d]$ linear code $C$ over $\mathbb{F}_p$ is a $k$-dimensional subspace of $\mathbb{F}_p^n$ with minimum distance $d$.  

We recall the definition of the generalized Hamming weights of a linear code \cite{wei1991generalized}.
Suppose that $U$ is an $r$-dimensional subspace of $C$,  the support of $U$ is defined to be $\text{Supp}(U)=\cup_{x \in U}\text{Supp}(x)$, where $\text{Supp}(x)$ is the set of coordinates where $x$ is nonzero, i.e.,
\[
 \text{Supp}(U) = \{i : 1 \le i \le n, \ x_i \neq 0 \text{ for some } x=(x_1,x_2,\ldots,x_n) \in U \}.
\]

\begin{df}
Let $C$ be an $[n,k,d]$ linear code over $\mathbb{F}_p$.
For $1 \le r \le k$, 
\[
d_r (C)=\min \{ \left |\text{Supp}(U) \right |: U \subset C, \  \text{dim } U=r\}
\]
is called the $r$-th \emph{generalized Hamming weight} (GHW) of $C$ and $\{d_r (C) : 1 \le r \le k\}$ is called the \emph{weight hierarchy} of $C$. 
\end{df}

The notion of generalized Hamming weights $d_r(C)$, which can be dated back to the 1970's  \cite{helleseth1977weight,klove1978weight},  is a natural generalization of the minimum distance $d=d_1(C)$.
In 1991, Wei first use it to characterize the cryptography performance of a linear code over the wire-tap channel of type II \cite{wei1991generalized}. 
Among other applications there are $t$-resilient functions \cite{tsfasman1995geometric},  trellis complexity of linear block codes \cite{forney1994dimension}, list decoding from erasures \cite{guruswami2003list} and secure network coding for wiretap networks  \cite{el2012secure,ngai2011network}.
A connection between GHWs and the covering radius of linear codes was found by Janwa and Lal \cite{janwa2007generalized}. 

For a set $D=\{d_1,d_2,\ldots,d_n \}\subset \mathbb{F}_q$, define a linear code of length $n$ over $\mathbb{F}_p$ by
\begin{equation}
C_D=\{c(a)=(\text{Tr}(a d_1), \text{Tr}(a d_2), \ldots, \text{Tr}(a d_n)): a \in \mathbb{F}_q\}. \label{cd}
\end{equation}
$D$ is called the \emph{defining set} of $C_D$.
The construction was first proposed by Ding et al. in \cite{ding2007cyclotomic} and many linear codes with few weights could be produced by chosing appropriate defining sets \cite{ding2015codes,ding2015linear,ding2007cyclotomic,ding2014binary,ding2015class}. 

In this paper, we shall employ Gauss periods to investigate the generalized Hamming weights of the following three classes of $C_D$:
\begin{align}
D &=\{\theta^i: i=0,1,\ldots, n_1-1\}, \label{ds1} \\
D &=\{\theta^i: i=0,1,\ldots, n_2-1\},  \text{ where } n_2=\frac{N_2}{N}, \label{ds2}\\
D &\text{ is a skew set of } \mathbb{F}_q. \label{ds3} 
\end{align}
When $D$ is given by \eqref{ds1} and \eqref{ds2} we present two general formulas on $d_r(C_D)$ involving Gauss periods. 
Using the results on Gauss periods we can determine $d_r(C_D)$ for small $r$ in the case $1 \le N_1 \le 4$ and in the semiprimitive case.
When $D$ is given by \eqref{ds3} we can determine the weight hierarchy of $C_D$.

It should be remarked that when $D$ is given by \eqref{ds1}, $C_D$ is in fact the irreducible cyclic code of length $n_1$ over $\mathbb{F}_p$. In \cite{yang2015generalized} Yang et al. consider the generalized Hamming weights of the irreducible cyclic code over $\mathbb{F}_q$. Essentially they have obtained our results in the case $N_1=1,2$ and in the semiprimitive case. But in the semiprimitive case their results are incomplete with a problem left. 
In this paper we solve this problem and extent the discussions to the case $1 \le N_1 \le 4$.

The rest of this paper is organized as follows. 
In Sect. \ref{secpr}, we introduce some notations and basic results on Gauss periods used in this paper. 
In Sect. \ref{secfi}, we investigate the generalized Hamming weights of three classes of $C_D$.
In Sect. \ref{secco}, we conclude this paper.

\section{Preliminaries} \label{secpr}
Firstly, we introduce several bounds on GHWs of a linear code.
\begin{lem}[\cite{tsfasman1995geometric}]
Let $C$ be an $[n,k]$ linear code over $\mathbb{F}_p$. 
For $1 \le r \le k$,
\begin{description}
\item[(1) Singleton type bound:]
\[
r \le d_r(C) \le n-k+r,
\]
\item[(2) Plotkin-like bound:]
\[
d_r(C) \le \left \lfloor \frac{n(p^r-1)p^{k-r}}{p^k-1} \right \rfloor, 
\] 
\item[(3) Griesmer-like bound:]
\[
d_r(C) \ge \sum_{i=0}^{r-1} \left \lceil \frac{d_1(C)}{p^i} \right \rceil.
\]
\end{description}
\end{lem}

The dimension of $C_D$ defined by \eqref{cd} is determined by 
$D$ and we have the following lemma:
\begin{lem}[\cite{ding2007cyclotomic}]  \label{dim}
Let $S_p(D)$ denote the $\mathbb{F}_p$-subspace of  $\mathbb{F}_q$ spanned by $D$, then
\[
\text{dim } C_D=\text{dim }S_p(D).
\]
\end{lem}

In the remaining parts of this section we introduce basic results on Gauss periods. For more details we refer to the book \cite{ding2015codes} and the paper \cite{myerson1981period}.
Let $\zeta_p=e^{\frac{2 \pi \sqrt{-1}}{p}}$ be the primitive $p$-th root of unity.
The canonical additive character of $\mathbb{F}_q$ can be defined by
\begin{align*}
\psi: \mathbb{F}_q &\rightarrow \mathbb{C}^*, \\
x &\mapsto \zeta_p^{\text{Tr}(x)}.
\end{align*}
The orthogonal property of additive characters is given by
\[
\sum_{x \in \mathbb{F}_q} \psi (ax)=\begin{cases}
q, & \text{if} \ a=0, \\
0, & \text{otherwise}.
\end{cases}
\]

Let $ \langle \theta \rangle$  denote the subgroup of the cyclic group $\mathbb{F}_q^*=\mathbb{F}_q \backslash \{0\}$ generated by $\theta$ and $C_i^{(N,q)}$ denote the $i$th left coset, namely $C_i^{(N,q)}=\alpha^i   \langle \theta  \rangle$. 
These cosets are called the \emph{cyclotomic classes} of order $N$ in  $\mathbb{F}_q$. 

\begin{df}
Suppose that $\psi$ is the canonical additive character of $\mathbb{F}_q$.
The  \emph{Gaussian periods} of order $N$ are defined by
\[
\eta_i^{(N,q)}=\sum_{x \in C_i^{(N,q)}} \psi (x), \ i=0,1,\ldots,N-1.
\]
\end{df}

\begin{lem} \label{l1}
\[
\left | \eta_i^{(N,q)}+\frac{1}{N} \right | \le \frac{(N-1)\sqrt{q}}{N}. 
\]
\end{lem}

The exact values of Gauss periods have been calculated for  several particular cases. 
To present known results we need to introduce the period polynomial which is defined to be 
\[
\Psi_{(N,q)}(X)=\prod_{i=0}^{N-1}(X-\eta_i^{(N,q)}).
\]

\begin{lem}  \label{l2}
Suppose that $q=p^m$ and $N=2$. 
The Gaussian periods are given by
\[
\eta_0^{(2,q)}=\begin{cases}
\frac{-1+(-1)^{m-1}\sqrt{q}}{2}, & \text{if} \ p \equiv 1 \pmod{4},\\
\frac{-1+(-1)^{m-1}(\sqrt{-1})^m \sqrt{q}}{2}, & \text{if} \ p \equiv 3 \pmod{4}
\end{cases}
\]
and
\[
\eta_1^{(2,q)}=-1-\eta_0^{(2,q)}.
\]
\end{lem}

\begin{lem}  \label{l3}
Suppose that $q=p^m$ and $N=3$. 
If $p \equiv 2 \pmod{3}$, then $m$ is even and the factorization of $\Psi_{(3,q)}(X)$ is given by 
\[
\Psi_{(3,q)}(X)=\begin{cases}
3^{-3}(3X+1+2\sqrt{q})(3X+1-\sqrt{q})^2, & \text{if} \ m \equiv 0 \pmod{4}, \\
3^{-3}(3X+1-2\sqrt{q})(3X+1+\sqrt{q})^2, & \text{if} \ m \equiv 2 \pmod{4}.
\end{cases}
\]
\end{lem}

\begin{lem} \label{l4}
Suppose that $q=p^m$ and $N=3$.  
If $p \equiv 1 \pmod{3}$ and $m \equiv 0 \pmod{3}$, then the factorization of $\Psi_{(3,q)}(X)$ is given by
\[
\Psi_{(3,q)}(X) =  3^{-3}(3X+1-c_1 q^{\frac{1}{3}})
\left(3X+1+\frac{1}{2}(c_1+9d_1)q^{\frac{1}{3}}\right) 
\left(3X+1+\frac{1}{2}(c_1-9d_1)q^{\frac{1}{3}}\right),
\]
where $c_1$ and $d_1$ are given by $4p^{\frac{m}{3}}=c_1^2+27d_1^2$, $c_1 \equiv 1 \pmod 3$ and $gcd(c_1,p)=1$.
These restrictions determine $c_1$ uniquely, and $d_1$ up to sign.
\end{lem}

\begin{lem} \label{l5}
Suppose that $q=p^m$ and $N=4$. 
If $p \equiv 3 \pmod{4}$, then $m$ is even and the factorization of $\Psi_{(4,q)}(X)$ is given by
\[
\Psi_{(4,q)}(X)=\begin{cases}
4^{-4}(4X+1+3\sqrt{q})(4X+1-\sqrt{q})^3, & \text{if} \ m \equiv 0 \pmod{4}, \\
4^{-4}(4X+1-3\sqrt{q})(4X+1+\sqrt{q})^3, & \text{if} \ m \equiv 2 \pmod{4}.
\end{cases}
\]
\end{lem}

\begin{lem} \label{l6}
Suppose that $q=p^m$ and $N=4$.  
If $p \equiv 1 \pmod{4}$ and $m \equiv 0 \pmod{4}$, then the factorization of $\Psi_{(4,q)}(X)$ is given by
\begin{align*}
\Psi_{(4,q)}(X)& =  4^{-4}((4X+1)+\sqrt{q}+2q^{\frac{1}{4}}u_1)((4X+1)+\sqrt{q}-2q^{\frac{1}{4}}u_1) \\
& \ \times ((4X+1)-\sqrt{q}+4q^{\frac{1}{4}}v_1)((4X+1)-\sqrt{q}-4q^{\frac{1}{4}}v_1),
\end{align*}
where $u_1$ and $v_1$ are given by $p^{\frac{m}{2}}=u_1^2+4v_1^2$, $u_1 \equiv 1 \pmod 4$ and $gcd(u_1,p)=1$.
These restrictions determine $u_1$ uniquely, and $v_1$ up to sign.
\end{lem}

\begin{lem}[The semiprimitive case] \label{l7}
Assume that $N \ge 3$ and $q=p^{2j\gamma}$, where $N | (p^j+1)$ and $j$ is the smallest such positive integer. Then the Gaussian periods of order $N$ are given below:
\begin{description}
\item[(a)] If $\gamma, p, \frac{p^j+1}{N}$ are all odd, then
\[
\eta_{\frac{N}{2}}^{(N,q)}=\sqrt{q}-\frac{\sqrt{q}+1}{N}, \
\eta_i^{(N,q)}=-\frac{\sqrt{q}+1}{N}  \text{ for all } i \neq \frac{N}{2}.
\]
\item[(b)] In all the other cases,
\[
\eta_0^{(N,q)}=\frac{(-1)^{\gamma+1}(N-1)\sqrt{q}-1}{N}, \
\eta_i^{(N,q)}=\frac{(-1)^{\gamma}\sqrt{q}-1}{N} \text{ for all } i \neq 0.
\]
\end{description}
\end{lem}

\section{Generalized Hamming weights of  $C_D$} \label{secfi}
In this section we investigate the generalized Hamming weights of the linear code $C_D$ defined by \eqref{cd}.
For convenience we denote by $\begin{bmatrix} r \\ U \end{bmatrix}_p$ the set of $r$-dimensional subspaces of $U$ for any $\mathbb{F}_p$-vector space $U$.

We begin to consider the $\mathbb{F}_p$-linear surjective mapping
\begin{align} 
c: \mathbb{F}_q & \rightarrow C_D, \notag \\
a & \mapsto c(a).  \label{c}
\end{align}
In this paper we need $c$ to be an isomorphism, which is automatic when $D$ is given by \eqref{ds1} and \eqref{ds3}.
When $D$ is given by \eqref{ds2}, we can prove that  $c$ is an isomorphism when $1 \le N_1 \le \sqrt{q}$ (see Theorem \ref{iso} below).

For $1 \le r \le m$, $c$ induces the following bijection:
\begin{align*}
c^r: \begin{bmatrix}  r  \\ \mathbb{F}_q  \end{bmatrix}_p & \rightarrow \begin{bmatrix} r \\ C_D \end{bmatrix}_p, \\
H^r & \mapsto U^r=\{c(a): a \in H^r\}.
\end{align*}
Set $n=|D|$, by definition
\begin{align}
d_r(C_D) &=\min \left\{|\text{Supp}(U^r)|: U^r \in \begin{bmatrix} r \\ C_D \end{bmatrix}_p \right\} \notag \\
&=n-\max \left\{ N(U^r):  U^r \in \begin{bmatrix} r \\ C_D \end{bmatrix}_p \right\},
\label{d2}
\end{align}
where
\begin{align*}
N(U^r)&= | \{i : x_i=0 \text{ for all } x=(x_1, x_2,\ldots, x_n) \in U^r\}|  \\
&=|\{i:\text{Tr}(a d_i)=0 \text{ for all } a \in H^r\}|  \\
&=\left| \left\{ i: \text{$\{a_1,\ldots, a_r\}$ is an $\mathbb{F}_p$-basis of $H^r$}, \ \text{Tr}(a_j d_i)=0 \text{ for all } j \right\} \right|. 
\end{align*}
By the orthogonal property of additive characters
\begin{align} 
N(U^r)&= \sum_{i=1}^{n} \prod_{j=1}^r \left( \frac{1}{p}\sum_{z_j \in \mathbb{F}_p} \zeta_p^{z_j \text{Tr}(a_j d_i)} \right) \notag \\
&=\frac{1}{p^r} \sum_{d \in D} \sum_{z_1,\ldots,z_r \in \mathbb{F}_p}\psi \left( (\sum_{j=1}^r z_j a_j)d \right) \notag \\
&=\frac{1}{p^r} \sum_{d \in D} \sum_{a \in H^r}\psi (a d) \notag \\
&=\frac{n}{p^r}+\frac{1}{p^r} \sum_{a \in H^r \backslash \{0\}} \sum_{d \in D} \psi (a d). 
\label{d3}
\end{align}

\subsection{$D$ given by \eqref{ds1}}
When $D$ is given by \eqref{ds1},  $S_p(D)=\mathbb{F}_p(\theta)=\mathbb{F}_q$.
By Lemma \ref{dim}, $C_D$ is an $[n_1, m]$ linear code and  $c$ in \eqref{c} is an isomorphism.

\begin{lem}[\cite{ding2015codes}] \label{mul}
Let $e$ be a positive divisor of $q-1$ and let $i$ be an integer with $0 \le i < e$. We have the following multiset equality:
\[
\left\{ xy: y \in \mathbb{F}_p^*, x \in C_i^{(e, q)}\right\}=\frac{(p-1) gcd(\frac{q-1}{p-1},e)}{e}*C_i^{(gcd(\frac{q-1}{p-1},e),q)},
\]
where $\frac{(p-1) gcd(\frac{q-1}{p-1},e)}{e}*C_i^{(gcd(\frac{q-1}{p-1},e),q)}$ denotes the multiset in which each element in the set $C_i^{(gcd(\frac{q-1}{p-1},e),q)}$ appears in the multiset with multiplicity $\frac{(p-1) gcd(\frac{q-1}{p-1},e)}{e}$.
\end{lem}

\begin{thm}
Let $D$ be given by \eqref{ds1} and $H^r_i=H^r \cap   C_i^{(N_1,q)}$ for $H^r \in \begin{bmatrix} r \\ \mathbb{F}_q \end{bmatrix}_p$.
For $1 \le r \le m$,
\begin{equation}
d_r(C_D) = n_1(1-\frac{1}{p^r})-\frac{N_1}{p^r N} \times  \max \left\{  \sum_{i=0}^{N_1-1} | H^r_i | \eta_i^{(N_1,q)}:  H^r \in \begin{bmatrix} r \\ \mathbb{F}_q \end{bmatrix}_p \right\}.
\label{ghw1}
\end{equation}
 \end{thm} 

\begin{pf}
Note that  the multiplication by any $z \in \mathbb{F}_p^*$ is a linear automorphism of $H^r$.
By formula \eqref{d3} and  Lemma \ref{mul} we have
\begin{align} 
N(U^r)&=\frac{n_1}{p^r}+\frac{1}{p^r(p-1)} \sum_{a \in H^r \backslash \{0\}} \sum_{z \in \mathbb{F}_p^*} \sum_{d \in D} \psi (za d) \notag \\
&=\frac{n_1}{p^r}+\frac{1}{p^r(p-1)} \sum_{a \in H^r \backslash \{0\}} \sum_{z \in \mathbb{F}_p^*} \sum_{i=0}^{n_1-1} \psi (za \theta^i) \notag \\
&=\frac{n_1}{p^r}+\frac{N_1}{p^r N} \sum_{a \in H^r \backslash \{0\}} \sum_{z \in C_0^{(N_1,q)}} \psi (a z) \notag \\
&=\frac{n_1}{p^r}+\frac{N_1}{p^r N} \sum_{i=0}^{N_1-1} | H^r_i | \eta_i^{(N_1,q)}.  
\label{d4}
\end{align}

The formula \eqref{ghw1} is derived from \eqref{d2}  and \eqref{d4}.
\end{pf}

\begin{cll} \label{c1}
Let $D$ be given by \eqref{ds1}.
If $N_1=1$, for $1 \le r \le m$,
\[
d_r(C_D) = (1-\frac{1}{p^r})\frac{q}{N}.
\]
\end{cll}

\begin{pf}
Since $C_0^{(1,q)}=\mathbb{F}_q^*$,  $\eta_0^{(1,q)}=\sum \limits_{x \in \mathbb{F}_q^*} \psi (x)=-1$ and $H_0^r=H^r \backslash \{0\}$.
By formula \eqref{ghw1}
\[
d_r(C_D) = n_1(1-\frac{1}{p^r})-\frac{(p^r-1)(-1)}{p^r N}=(1-\frac{1}{p^r})\frac{q}{N}.
\]
\end{pf}

If $N_1 \ge 2$, let $m_1$ be a positive factor of $m$ such that $N_1 \left| \frac{p^m-1}{p^{m_1}-1} \right.$, then  $\mathbb{F}_{p^{m_1}}^* \subset C_0^{(N_1,q)}$.
Without loss of generality we assume that $\max \left\{\eta_i^{(N_1,q)}: i=0,1,\ldots, N_1-1 \right\}=\eta_0^{(N_1,q)}$.
For $1 \le r \le m_1$, we can take $H^r$ to be an r-dimensional subspace of $\mathbb{F}_{p^{m_1}}$ so that $H_0^r=H^r \backslash \{0\}$ and $H_i^r=\varnothing$ for all $i \ne 0$.
Therefore the maximal value in right-hand side of \eqref{ghw1} is
$(p^r-1)\max \left\{\eta_i^{(N_1,q)}: i=0,1,\ldots, N_1-1 \right\}$.
Using lemmas on Gauss periods in Sect. \ref{secpr} we can easily get the corollaries in the remaining parts of this subsection.

\begin{cll} 
Let $D$ be given by \eqref{ds1}.
If $N_1=2$, for $1 \le r \le \frac{m}{2}$,
\[
d_r(C_D) = (1-\frac{1}{p^r})\frac{q-\sqrt{q}}{N}.
\]
\end{cll}

\begin{pf}
Since $N_1=gcd(N,\frac{q-1}{p-1})=2$, $m$ is even and $2 \left| \frac{p^m-1}{p^{m/2}-1} \right.$.
By Lemma \ref{l2} 
\[
\max \left\{\eta_0^{(2,q)}, \eta_1^{(2,q)} \right\}=\frac{-1+ \sqrt{q}}{2}.
\]
So for $1 \le r \le \frac{m}{2}$, 
\[
d_r(C_D) = n_1(1-\frac{1}{p^r})-\frac{(-1+ \sqrt{q})(p^r-1)}{p^r N}= (1-\frac{1}{p^r})\frac{q-\sqrt{q}}{N}.
\]
\end{pf}

\begin{cll}
Let $D$ be given by \eqref{ds1}.
If $N_1=3$, $p \equiv 2 \pmod{3}$ and $m \equiv 2 \pmod{4}$, for $1 \le r \le \frac{m}{2}$,
\[
d_r(C_D) =(1-\frac{1}{p^r})\frac{q-2\sqrt{q}}{N}.
\]
\end{cll}

\begin{pf}
Since $p \equiv 2 \pmod{3}$ and $m \equiv 2 \pmod{4}$,  $3 \left| \frac{p^m-1}{p^{m/2}-1} \right.$.
By Lemma \ref{l3}
\[
\max \left\{\eta_0^{(3,q)}, \eta_1^{(3,q)}, \eta_2^{(3,q)} \right\}=\frac{-1+ 2\sqrt{q}}{3}.
\]
So for $1 \le r \le \frac{m}{2}$, 
\[
d_r(C_D) = n_1(1-\frac{1}{p^r})-\frac{(-1+ 2\sqrt{q})(p^r-1)}{p^r N}= (1-\frac{1}{p^r})\frac{q-2\sqrt{q}}{N}.
\]

\end{pf}

\begin{cll}
Let $D$ be given by \eqref{ds1}.
If $N_1=3$, $p \equiv 2 \pmod{3}$ and $4l |m$ where $l$ is an odd positive integer, for $1 \le r \le l$,
\[
d_r(C_D) =(1-\frac{1}{p^r})\frac{q-\sqrt{q}}{N}.
\]
\end{cll}
\begin{pf}
Since $p \equiv 2 \pmod{3}$ and $l$ is odd,  $3 \left| \frac{p^m-1}{p^l-1} \right.$.
By Lemma \ref{l3}
\[
\max \left\{\eta_0^{(3,q)}, \eta_1^{(3,q)}, \eta_2^{(3,q)} \right\}=\frac{-1+ \sqrt{q}}{3}.
\]
So for $1 \le r \le l$, 
\[
d_r(C_D) = n_1(1-\frac{1}{p^r})-\frac{(-1+ \sqrt{q})(p^r-1)}{p^r N}= (1-\frac{1}{p^r})\frac{q-\sqrt{q}}{N}.
\]
\end{pf}

\begin{cll} 
Let $D$ be given by \eqref{ds1}.
If $N_1=3$ and $p \equiv 1 \pmod{3}$, for $1 \le r \le \frac{m}{3}$,
\begin{equation} 
d_r(C_D) = \begin{cases}
(1-\frac{1}{p^r})\frac{q-c_1 q^{\frac{1}{3}}}{N}, & \text{if} \ c_1>3|d_1|,\\
(1-\frac{1}{p^r})\frac{q+\frac{1}{2}(c_1-9|d_1|) q^{\frac{1}{3}}}{N}, & \text{if} \ c_1<3|d_1|, 
 \end{cases}
\end{equation}
where $c_1,d_1$ are defined in Lemma \ref{l4}.
\end{cll}

\begin{pf}
Since $N_1=3$ and $p \equiv 1 \pmod{3}$, $m \equiv 0 \pmod{3}$ and $3 \left| \frac{p^m-1}{p^{m/3}-1} \right.$.
By Lemma \ref{l4}
\[
\max \left\{\eta_0^{(3,q)}, \eta_1^{(3,q)}, \eta_2^{(3,q)} \right\}=\begin{cases}
\frac{-1+c_1 q^{\frac{1}{3}}}{3} &  \text{if}  \  c_1>3|d_1|, \\
\frac{-1-\frac{1}{2}(c_1-9|d_1|) q^{\frac{1}{3}}}{3} &  \text{if}  \  c_1<3|d_1|.
\end{cases}
\]
\end{pf}

\begin{cll}
Let $D$ be given by \eqref{ds1}.
If $N_1=4$, $p \equiv 3 \pmod{4}$ and $m \equiv 2 \pmod{4}$, for $1 \le r \le \frac{m}{2}$,
\[
d_r(C_D) =(1-\frac{1}{p^r})\frac{q-3\sqrt{q}}{N}.
\]
\end{cll}

\begin{pf}
Since $p \equiv 3 \pmod{4}$ and $m \equiv 2 \pmod{4}$,  $4 \left| \frac{p^m-1}{p^{m/2}-1} \right.$.
By Lemma \ref{l5}
\[
\max \left\{\eta_0^{(4,q)}, \eta_1^{(4,q)}, \eta_2^{(4,q)}, \eta_3^{(4,q)} \right\}=\frac{-1+ 3\sqrt{q}}{4}.
\]
\end{pf}

\begin{cll}
Let $D$ be given by \eqref{ds1}.
If $N_1=4$, $p \equiv 3 \pmod{4}$ and $m \equiv 0 \pmod{4}$, for $1 \le r \le \frac{m}{4}$,
\[
d_r(C_D) =(1-\frac{1}{p^r})\frac{q-\sqrt{q}}{N}.
\]
\end{cll}

\begin{pf}
Since $p \equiv 3 \pmod{4}$ and $m \equiv 0 \pmod{4}$,  $4 \left| \frac{p^m-1}{p^{m/4}-1} \right.$.
By Lemma \ref{l5}
\[
\max \left\{\eta_0^{(4,q)}, \eta_1^{(4,q)}, \eta_2^{(4,q)}, \eta_3^{(4,q)} \right\}=\frac{-1+ \sqrt{q}}{4}.
\]
\end{pf}

\begin{cll}
Let $D$ be given by \eqref{ds1}.
If $N_1=4$ and $p \equiv 1 \pmod{4}$,  for $1 \le r \le \frac{m}{4}$,
\begin{equation} 
d_r(C_D) =
(1-\frac{1}{p^r})\frac{q-\sqrt{q}-4q^{\frac{1}{4}}|v_1|}{N},
 \end{equation}
 where $u_1,v_1$ are defined in Lemma \ref{l6}.
\end{cll}

\begin{pf}
Since $N_1=4$ and $p \equiv 1 \pmod{4}$, $m \equiv 0 \pmod{4}$ and $4 \left| \frac{p^m-1}{p^{m/4}-1} \right.$.
By Lemma \ref{l6}
\[
\max \left\{\eta_0^{(4,q)}, \eta_1^{(4,q)}, \eta_2^{(4,q)}, \eta_3^{(4,q)} \right\}=\frac{-1+ \sqrt{q}+4q^{\frac{1}{4}}|v_1|}{4}.
\]
\end{pf}

\begin{cll}
Let $D$ be given by \eqref{ds1}.
Assume that $N_1 \ge 3$ and $q=p^{2j\gamma}$, where $N_1 | (p^j+1)$ and $j$ is the smallest such positive integer. 
For $1 \le r \le j$,
\[ 
d_r(C_D) = \begin{cases}
(1-\frac{1}{p^r})\frac{q-\sqrt{q}}{N}, &\text{if} \  \gamma \text{ is even},\\
(1-\frac{1}{p^r})\frac{q-(N_1-1)\sqrt{q}}{N}, &\text{if} \  \gamma \text{ is odd}.
\end{cases}
\]
\end{cll}

\begin{pf}
Since $q=p^{2j\gamma}$, $m=2j\gamma$ and
\[
\frac{p^m-1}{p^j-1}=\sum_{i=0}^{2\gamma-1}(p^j)^i
\equiv \sum_{i=0}^{2\gamma-1}(-1)^i \equiv 0 \pmod {N_1}.
\]
By Lemma \ref{l7}
\[
\max \left\{\eta_i^{(N_1,q)}: i=0,1,\ldots, N_1-1 \right\}= \begin{cases}
\frac{\sqrt{q}-1}{N}, &\text{if} \  \gamma \text{ is even},\\
\frac{(N_1-1)\sqrt{q}-1}{N}, &\text{if} \  \gamma \text{ is odd}.
\end{cases}
\]
\end{pf}

\begin{rk}
It's easy to check all computed values of $d_r(C_D)$ in the above  corollaries meet the Griesmer-like bound. Moreover by Corollary \ref{c1}, $d_r(C_D)$ also meets the Plotkin-like bound for all $r$ if $N_1=1$.
\end{rk}

\begin{rk}
The value of $d_r(C_D)$ in the semiprimitive case discussed in \cite{yang2015generalized} is a little complicated,  where the authors considered two subcases. Essentially they failed to compute $d_r(C_D)$ when $\gamma$ is odd and $N_1$ is even (See the remark after Theorem 13 in \cite{yang2015generalized}). 
In this paper we solve this problem.
\end{rk}

\begin{rk}
By considering the orthogonal complementary space of $H^r$ for $H^r \in \begin{bmatrix} r \\ \mathbb{F}_q \end{bmatrix}_p$
and using techniques developed in \cite{yang2015generalized} we can determine $d_r(C_D)$ for $m-m_1 \le r \le m$, where $m_1$ is a positive factor of $m$ with $N_1 \left| \frac{p^m-1}{p^{m_1}-1} \right.$.
Particularly when $m_1=\frac{m}{2}$ (e.g. $N_1=2$), the weight hierarchy of $C_D$ can be determined.
For details the readers are referred to that paper.
\end{rk}
 
\subsection{$D$ given by \eqref{ds2}}
When $D$ is given by \eqref{ds2},  $C_D$ is an $[n_2, m]$ linear code and  $c$ in \eqref{c} is an isomorphism when $1 \le N_1 \le \sqrt{q}$ by the following theorem:

\begin{thm} \label{iso}
Let $D$ be given by \eqref{ds2}. 
If $1 \le N_1 \le \sqrt{q}$, the Hamming weight of the codeword $c(a)$ is nonzero for $a \in \mathbb{F}_q^*$.
\end{thm}
\begin{pf}
We need a naive theorem in group theory:

\vspace{.5em}
\begin{minipage}{.97\linewidth}
Suppose that $H$ and $K$ are two subgroups of a group $G$. For $h_1,h_2 \in H$, $h_1K=h_2K$ holds iff $h_1(H \cap K) =h_2(H \cap K)$ holds. Moreover, $HK/K \cong H/(H \cap K)$ where $HK=\{hk:h \in H, k \in K\}$.
\end{minipage}
\vspace{.5em}

\noindent Take $H=C_0^{(N,q)}$ and $K=\mathbb{F}_p^*$,
then $HK=C_0^{(N_1,q)}$ and $H \cap K=C_0^{(N_2,q)}$. There is a coset decomposition of $H$:
\[
H=\bigcup_{i=0}^{n_2-1} \theta^i (H \cap K).
\]
So $D=\{\theta^i: i=0,1,\ldots, n_2-1\}$ is a complete set of coset representatives of $HK/K$.
When $a \in C_i^{(N_1,q)}$, $0 \le i < N_1$, the Hamming weight of $c(a)$ is
\begin{align*}
W_H(c(a)) &=n_2-\sum_{i=0}^{n_2-1} \left( \frac{1}{p}\sum_{z \in \mathbb{F}_p} \zeta_p^{z\text{Tr}(a\theta^i)} \right) \notag \\
&=n_2(1-\frac{1}{p})- \frac{1}{p}\sum_{i=0}^{n_2-1}\sum_{z \in \mathbb{F}_p^*}\psi(za \theta^i)\\
&=n_2(1-\frac{1}{p})- \frac{1}{p}\sum_{z \in C_0^{(N_1,q)}}\psi(az) \\
&=\frac{q}{pN_1}- \frac{1}{p}(\eta_i^{(N_1,q)}+\frac{1}{N_1}).
\end{align*}
If $1 \le N_1 \le \sqrt{q}$, by Lemma \ref{l1}
\[
W_H(c(a)) \ge \frac{q-(N_1-1)\sqrt{q}}{pN_1}>0. 
\] 
\end{pf}

\begin{thm}
Let $D$ be given by \eqref{ds2} with $1 \le N_1 \le \sqrt{q}$ and $H^r_i=H^r \cap   C_i^{(N_1,q)}$ for $H^r \in \begin{bmatrix} r \\ \mathbb{F}_q \end{bmatrix}_p$.
For $1 \le r \le m$,
\begin{equation}
d_r(C_D) = n_2(1-\frac{1}{p^r})-\frac{1}{p^r (p-1)} \times  \max \left\{  \sum_{i=0}^{N_1-1} | H^r_i | \eta_i^{(N_1,q)}:  H^r \in \begin{bmatrix} r \\ \mathbb{F}_q \end{bmatrix}_p \right\}.
\label{ghw2}
\end{equation}
\end{thm}

\begin{pf}
By formula \eqref{d3}  we have
\begin{align} 
N(U^r) &=\frac{n_2}{p^r}+\frac{1}{p^r(p-1)} \sum_{a \in H^r \backslash \{0\}} \sum_{z \in \mathbb{F}_p^*} \sum_{i=0}^{n_2-1} \psi (za \theta^i) \notag \\
&=\frac{n_2}{p^r}+\frac{1}{p^r(p-1)} \sum_{a \in H^r \backslash \{0\}} \sum_{z \in C_0^{(N_1,q)}} \psi (a z) \notag \\
&=\frac{n_2}{p^r}+\frac{1}{p^r(p-1)} \sum_{i=0}^{N_1-1} | H^r_i | \eta_i^{(N_1,q)}.  
\label{f2}
\end{align}

The formula \eqref{ghw2} is derived from \eqref{d2}  and \eqref{f2}.
\end{pf}

\begin{rk}
Similar corollaries can be obtained using the method in the above subsection, so we omit here.
\end{rk}

\subsection{$D$ given by \eqref{ds3}} 
\begin{df}
Let $D$ be a subset of $\mathbb{F}_q^*$. 
If $D$, $-D=\{-d: d \in D\}$ and $\{0\}$ form a partition of $\mathbb{F}_q$, then $D$ is called a \emph{skew set} of $\mathbb{F}_q$.  
\end{df}

When $D$ is given by \eqref{ds3},  the length of $C_D$ is $n_3=\frac{q-1}{2}$.
By Lemma \ref{dim}, the dimension of $C_D$ is $m$ noting that $|S_p(D)| \ge |D|$. 
So $c$ in \eqref{c} is an isomorphism. 

\begin{thm} \label{th}
Let $D$ be given by \eqref{ds3}.
For $1 \le r \le m$,
\begin{equation}
d_r(C_D)=(1-\frac{1}{p^r})\frac{q}{2}.
 \label{ghw3}
\end{equation}
\end{thm}

\begin{pf}
Note that $a D=\{a d: d \in D\}$ is also a skew set of $\mathbb{F}_q$ for any $a \in \mathbb{F}_q^*$.
By formula \eqref{d3}  we have
\begin{align}
N(U^r) &=\frac{n_3}{p^r}+\frac{1}{2p^r} \sum_{a \in H^r \backslash \{0\}}\left( \sum_{d \in D} \psi (a d)+ \sum_{d \in D} \psi (-a d) \right) \notag\\
&=\frac{n_3}{p^r}+\frac{1}{2p^r} \sum_{a \in H^r \backslash \{0\}} \sum_{d \in \mathbb{F}_q^*} \psi (d) \notag\\
&=\frac{q-1}{2p^r}+\frac{p^r-1}{2p^r}(-1) \notag \\
&= \frac{q-p^r}{2p^r}. \label{sk}
\end{align}

The formula \eqref{ghw3} is derived from \eqref{d2} and \eqref{sk}.
\end{pf}

For example, if $q \equiv 3 \pmod 4$, $ \langle \alpha^2 \rangle$ is a skew set of $\mathbb{F}_q$ and $gcd(2,\frac{q-1}{p-1})=1$. In this case the formulas of Corollary \ref{c1} and Theorem \ref{th} are in accordance.

\begin{rk}
By formula \eqref{ghw3} it's easy to check that $d_r(C_D)$ meets the Plotkin-like bound and the Griesmer-like bound for all $r$.
\end{rk}

\section{Concluding remarks} \label{secco}
In this paper we investigate the generalized Hamming weights of three classes of linear codes and determine them partly for some cases.
The generalized Hamming weights of many families of codes have been determined, such as  Hamming codes \cite{wei1991generalized}, Reed-Muller codes \cite{heijnen1998generalized}, BCH codes \cite{geer1994generalized,cheng1997generalized,van1996note}, cyclic codes \cite{feng1992generalized,janwa1997generalized}, trace codes \cite{stichtenoth1994generalized}, binary Kasami codes \cite{helleseth1995weight}, Melas and dual Melas codes \cite{van1994generalized},  AG codes \cite{barbero2000weight,yang1994weight,munuera1994generalized,homma2009second,hirschfeld1994weight}, etc. 
The readers are referred to \cite{tsfasman1995geometric} for a survey.

\bibliography{thesis}  
\end{document}